\documentclass[letterpaper]{article} 
\usepackage{aaai18}  
\usepackage{times}  
\usepackage{helvet}  
\usepackage{courier}  
\usepackage{url}  
\usepackage{graphicx}  
\usepackage[title]{appendix}
\usepackage{subcaption}
\usepackage{multirow}
\usepackage{makecell}
\usepackage{amsmath}
\usepackage{arydshln}
\graphicspath{{figs/}}

\nocopyright

\def\z{\mathbf{z}}
\def\x{\mathbf{x}}
\def\y{\mathbf{y}}
\def\E{\mathbf{E}}

\def\t{\text{temp}}
\def\b{\text{bar}}

\begin{document}


\title{MuseGAN: Multi-track Sequential Generative Adversarial Networks for\\Symbolic Music Generation and Accompaniment}

\author{
Hao-Wen Dong\thanks{These authors contributed equally to this work.},\textsuperscript{1}
Wen-Yi Hsiao\footnotemark[1],\textsuperscript{1,2}
Li-Chia Yang,\textsuperscript{1}
Yi-Hsuan Yang\textsuperscript{1} \\
\textsuperscript{1}Research Center for Information Technology Innovation, Academia Sinica, Taipei, Taiwan \\
\textsuperscript{2}Department of Computer Science, National Tsing Hua University, Hsinchu, Taiwan \\
salu133445@citi.sinica.edu.tw,
s105062581@m105.nthu.edu.tw,
\{richard40148, yang\}@citi.sinica.edu.tw \\
}
\maketitle


\begin{abstract}
Generating music has a few notable differences from generating images and videos.
First, music is an art of time, necessitating a temporal model. Second, music is usually composed of multiple instruments/tracks with their own temporal dynamics, but collectively they unfold over time interdependently. Lastly, musical notes are often grouped into chords, arpeggios or melodies in polyphonic music, and thereby introducing a chronological ordering of notes is not naturally suitable. In this paper, we propose three models for symbolic multi-track music generation under the framework of generative adversarial networks (GANs). The three models, which differ in the underlying assumptions and accordingly the network architectures, are referred to as the jamming model, the composer model and the hybrid model. We trained the proposed models on a dataset of over one hundred thousand bars of rock music and applied them to generate piano-rolls of five tracks: bass, drums, guitar, piano and strings. A few intra-track and inter-track objective metrics are also proposed to evaluate the generative results, in addition to a subjective user study. We show that our models can generate coherent music of four bars right from scratch (i.e. without human inputs). We also extend our models to human-AI cooperative music generation: given a specific track composed by human, we can generate four additional tracks to accompany it. All code, the dataset and the rendered audio samples are available at \url{https://salu133445.github.io/musegan/}.
\end{abstract}


\section{Introduction}
\label{section:intro}

Generating realistic and aesthetic pieces has been considered as one of the most exciting tasks in the field of AI. Recent years have seen major progress in generating images, videos and text, notably using generative adversarial networks (GANs) \cite{GAN,DCGAN,VGAN,TGAN,SeqGAN}. Similar attempts have also been made to generate symbolic music, but the task remains challenging for the following reasons.

\begin{figure}[t]
  \centering
  \includegraphics[width=0.78\linewidth]{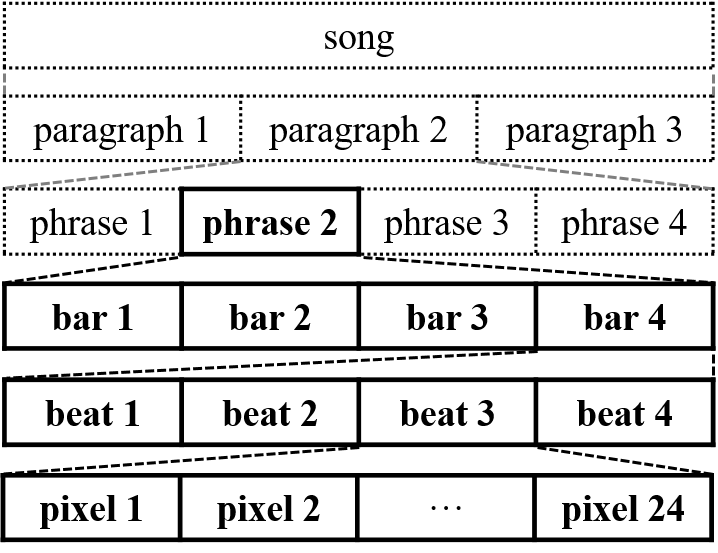}
  \caption{Hierarchical structure of a music piece.}
  \label{fig:hier}
\end{figure}

First, music is an art of time. As shown in Figure \ref{fig:hier}, music has a hierarchical structure, with higher-level building blocks (e.g., a phrase) made up of smaller recurrent patterns (e.g., a bar). People pay attention to structural patterns related to coherence, rhythm, tension and the emotion flow while listening to music \cite{herremans17tac}. Thus, a mechanism to account for the temporal structure is critical.

Second, music is usually composed of multiple instruments/tracks. A modern orchestra usually contains four different sections: brass, strings, woodwinds and percussion; a rock band often includes a bass, a drum set, guitars and possibly a vocal. These tracks interact with one another closely and unfold over time interdependently. In music theory, we can also find extensive discussions on composition disciplines for relating sounds, e.g., harmony and counterpoint.

Lastly, musical notes are often grouped into chords, arpeggios or melodies. It is not naturally suitable to introduce a chronological ordering of notes for polyphonic music. Therefore, success in natural language generation and monophonic music generation may not be readily generalizable to polyphonic music generation.

As a result, most prior arts (see the Related Work section for a brief survey) chose to simplify symbolic music generation in certain ways to render the problem manageable. Such simplifications include: generating only single-track monophonic music, introducing a chronological ordering of notes for polyphonic music, generating polyphonic music as a combination of several monophonic melodies, etc.

It is our goal to avoid as much as possible such simplifications. In essence, we aim to generate multi-track polyphonic music with 1) harmonic and rhythmic structure, 2) multi-track interdependency, and 3) temporal structure.

To incorporate a temporal model, we propose two approaches for different scenarios: one generates music from scratch (i.e. without human inputs) while the other learns to follow the underlying temporal structure of a track given \emph{a priori} by human. To handle the interactions among tracks, we propose three methods based on our understanding of how pop music is composed: one generates tracks independently by their \emph{private} generators (one for each); another generates all tracks jointly with only one generator; the other generates each track by its private generator with additional \emph{shared} inputs among tracks, which is expected to guide the tracks to be collectively harmonious and coordinated. To cope with the grouping of notes, we view bars instead of notes as the basic compositional unit and generate music one bar after another using transposed convolutional neural networks (CNNs), which is known to be good at finding local, translation-invariant patterns.

We further propose a few intra-track and inter-track objective measures and use them to monitor the learning process and to evaluate the generated results of different proposed models quantitatively. We also report a user study involving 144 listeners for a subjective evaluation of the results.

We dub our model as the \underline{mu}lti-track \underline{se}quential \underline{g}enerative \underline{a}dversarial \underline{n}etwork, or MuseGAN for short. Although we focus on music generation in this paper, the design is fairly generic and we hope it will be adapted to generate multi-track sequences in other domains as well.

Our contributions are as follows:
\begin{itemize}
\item We propose a novel GAN-based model for multi-track sequence generation.
\item We apply the proposed model to generate symbolic music, which represents, to the best of our knowledge, the first model that can generate multi-track, polyphonic music.
\item We extend the proposed model to track-conditional generation, which can be applied to human-AI cooperative music generation, or music accompaniment.
\item We present the Lakh Pianoroll Dataset (LPD), which contains 173,997 unique multi-track piano-rolls derived from the Lakh Midi Dataset (LMD)~\cite{raffel16phd}.
\item We propose a few intra-track and inter-track objective metrics for evaluating artificial symbolic music.
\end{itemize}

All code, the dataset and the rendered audio samples can be found on our project website.\footnote{\url{https://salu133445.github.io/musegan/}}


\section{Generative Adversarial Networks}
\label{section:background}

The core concept of GANs is to achieve adversarial learning by constructing two networks: the \emph{generator} and the \emph{discriminator} \cite{GAN}. The generator maps a random noise $\z$ sampled from a prior distribution to the data space. The discriminator is trained to distinguish real data from those generated by the generator, whereas the generator is trained to fool the discriminator. The training procedure can be formally modeled as a two-player minimax game between the generator $G$ and the discriminator $D$: 
\begin{equation}
\label{eq:gan}
\small
\min_{G} \max_{D} \E_{\x\sim p_d}[\log(D(\x))] + \E_{\z\sim p_z}[1-\log(D(G(\z)))] \,,
\end{equation}
where $p_d$ and $p_z$ represent the distribution of real data and the prior distribution of $\z$, respectively.

In a follow-up research \cite{WGAN}, they argue that using the Wasserstein distance, or the Earth Mover’s distance, instead of the Jensen-Shannon divergence used in the original formulation, can stabilize the training process and avoid mode collapsing. To enforce a $K$-Lipschitz constraint, weight clipping is used in Wasserstein GAN, while it is later on found to cause optimization difficulties. An additional \emph{gradient penalty} term for the objective function of the discriminator is then proposed in \cite{ImprovedWGAN}. The objective function of $D$ becomes
\begin{equation}
\small
\label{eq:wgan-gp}
\E_{\x\sim p_d}[D(\x)] - \E_{\z\sim p_z}[D(G(\z))] + \E_{\hat{\x}\sim p_{\hat{\x}}}[(\nabla_{\hat{\x}}\|\hat{\x}\|-1)^2] \,,
\end{equation}
where $p_{\hat{\x}}$ is defined sampling uniformly along straight lines between pairs of points sampled from $p_d$ and $p_g$, the model distribution. The resulting WGAN-GP model is found to have faster convergence to better optima and require less parameters tuning. Hence, we resort to the WGAN-GP model as our generative model in this work.

\begin{figure}[t]
\centering
\begin{subfigure}[]{.12\linewidth}
  \centering
  bass \\ [0.34cm]
  drums \\ [0.34cm]
  guitar \\ [0.34cm]
  strings \\ [0.34cm]
  piano \\ \vspace{0.45cm}
\end{subfigure}
\hspace{0.05cm}
\begin{subfigure}[]{0.4\linewidth}
  \centering
  \includegraphics[scale=0.4]{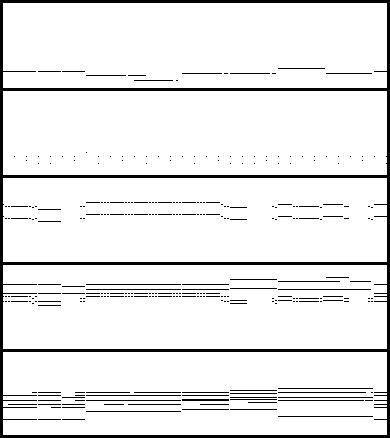} \\
  (a)
\end{subfigure}
\hfill
\begin{subfigure}[]{0.4\linewidth}
  \centering
  \includegraphics[scale=0.4]{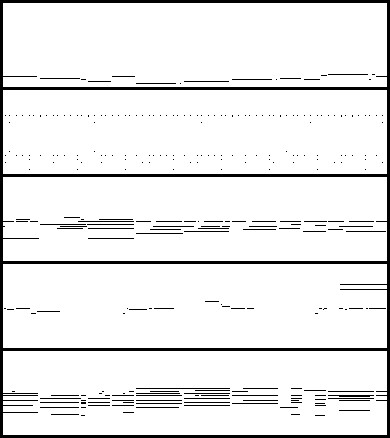} \\
  (b)
\end{subfigure}
\caption{Multi-track piano-roll representations of two music fragments of four bars with five tracks. The horizontal axis represents time, and the vertical axis represents notes (from low-pitched to high-pitched ones). A black pixel indicates that a specific note is played at that time step.}
\label{fig:train}
\end{figure}


\section{Proposed Model}
\label{section:model}

Following \cite{midinet}, we consider bars as the basic compositional unit for the fact that harmonic changes (e.g., chord changes) usually occur at the boundaries of bars and that human beings often use bars as the building blocks when composing songs.

\subsection{Data Representation}
\label{sec:data}

To model multi-track, polyphonic music, we propose to use the \emph{multiple-track piano-roll} representation. As exemplified in Figure~\ref{fig:train}, a piano-roll representation is a binary-valued, scoresheet-like matrix representing the presence of notes over different time steps, and a multiple-track piano-roll is defined as a set of piano-rolls of different tracks.

Formally, an $M$-track piano-roll of one bar is represented as a tensor $\x \in \{0, 1\}^{R \times S \times M}$, where $R$ and $S$ denote the number of time steps in a bar and the number of note candidates respectively. An $M$-track piano-roll of $T$ bars is represented as $\overrightarrow{\x} = \{\overrightarrow{\x}^{(t)}\}_{t=1}^T$, where $\overrightarrow{\x}^{(t)} \in \{0, 1\}^{R \times S \times M}$ denotes the multi-track piano-roll of bar $t$.

Note that the piano-roll of each bar, each track, for both the real and the generated data, is represented as a fixed-size matrix, which makes the use of CNNs feasible.

\subsection{Modeling the Multi-track Interdependency}
\label{sec:multitrack}

In our experience, there are two common ways to create music. Given a group of musicians playing different instruments, they can create music by improvising music without a predefined arrangement, a.k.a. jamming. Or, we can have a composer who arranges instruments with knowledge of harmonic structure and instrumentation. Musicians will then follow the composition and play the music. We design three models corresponding to these compositional approaches.

\subsubsection{Jamming Model}
\begin{figure}[t]
\centering
\includegraphics[width=.7\linewidth]{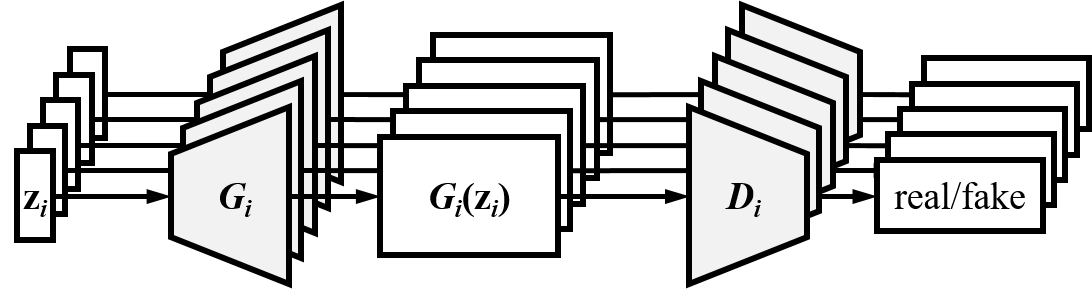}\\
(a) Jamming model \\
\vspace{0.2cm}
\includegraphics[width=.7\linewidth]{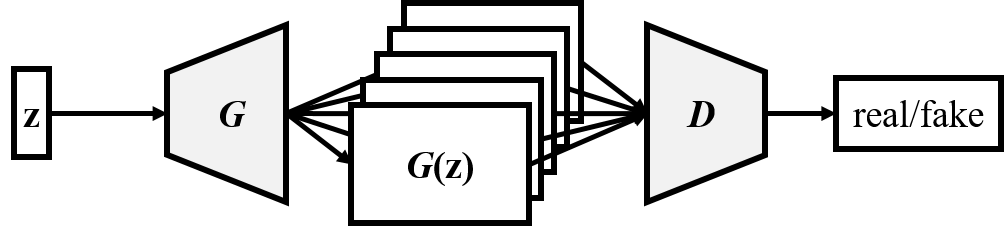}\\
(b) Composer model \\
\includegraphics[width=.7\linewidth]{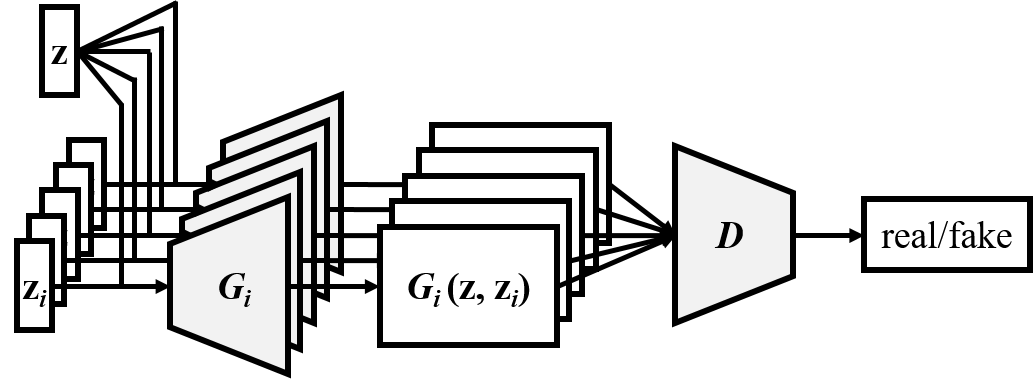}\\
(c) Hybrid model
\caption{Three GAN models for generating multi-track data. Note that we do not show the real data $\mathbf{x}$, which will also be fed to the discriminator(s).} 
\label{fig:multitrack}
\end{figure}

Multiple generators work independently and generate music of its own track from a \emph{private} random vector $\z_i$, $i = 1,2,\dots,M$, where $M$ denotes the number of generators (or tracks). These generators receive critics (i.e. backpropogated supervisory signals) from different discriminators. As illustrated in Figure~\ref{fig:multitrack}(a), to generate music of $M$ tracks, we need $M$ generators and $M$ discriminators.

\subsubsection{Composer Model}
One single generator creates a multi-channel piano-roll, with each channel representing a specific track, as shown in Figure~\ref{fig:multitrack}(b). This model requires only one \emph{shared} random vector $\z$ (which may be viewed as the intention of the composer) and one discriminator, which examines the $M$ tracks collectively to tell whether the input music is real or fake. Regardless of the value of $M$, we always need only one generator and one discriminator.

\subsubsection{Hybrid Model}
Combining the idea of jamming and composing, we further propose the hybrid model. As illustrated in Figure~\ref{fig:multitrack}(c), each of the $M$ generators takes as inputs an \emph{inter-track} random vector $\z$ and an \emph{intra-track} random vector $\z_i$. We expect that the inter-track random vector can coordinate the generation of different musicians, namely $G_i$, just like a composer does. Moreover, we use only one discriminator to evaluate the $M$ tracks collectively. That is to say, we need $M$ generators and only one discriminator.

A major difference between the composer model and the hybrid model lies in the flexibility---in the hybrid model we can use different network architectures (e.g., number of layers, filter size) and different inputs for the $M$ generators. Therefore, we can for example vary the generation of one specific track without losing the inter-track interdependency.

\begin{figure}[t]
\centering
\includegraphics[width=\linewidth]{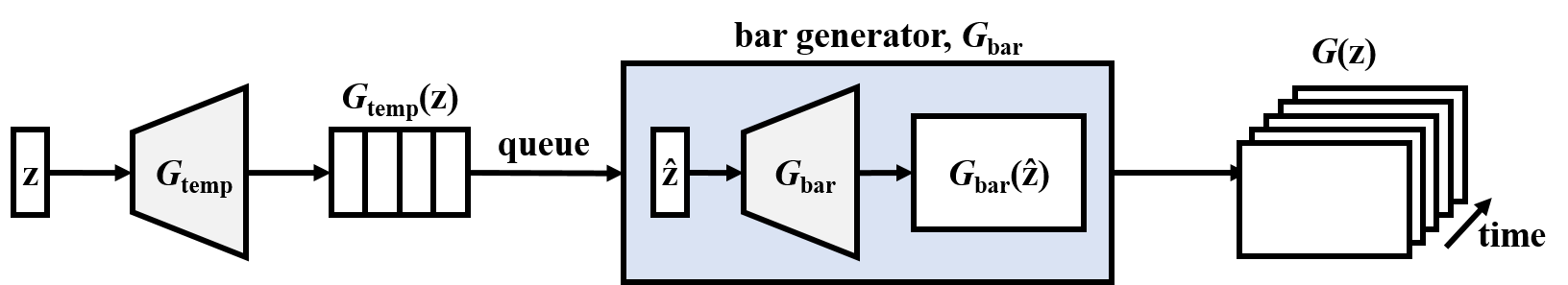} \\
(a) Generation from scratch \\
\vspace{0.2cm}
\includegraphics[width=\linewidth]{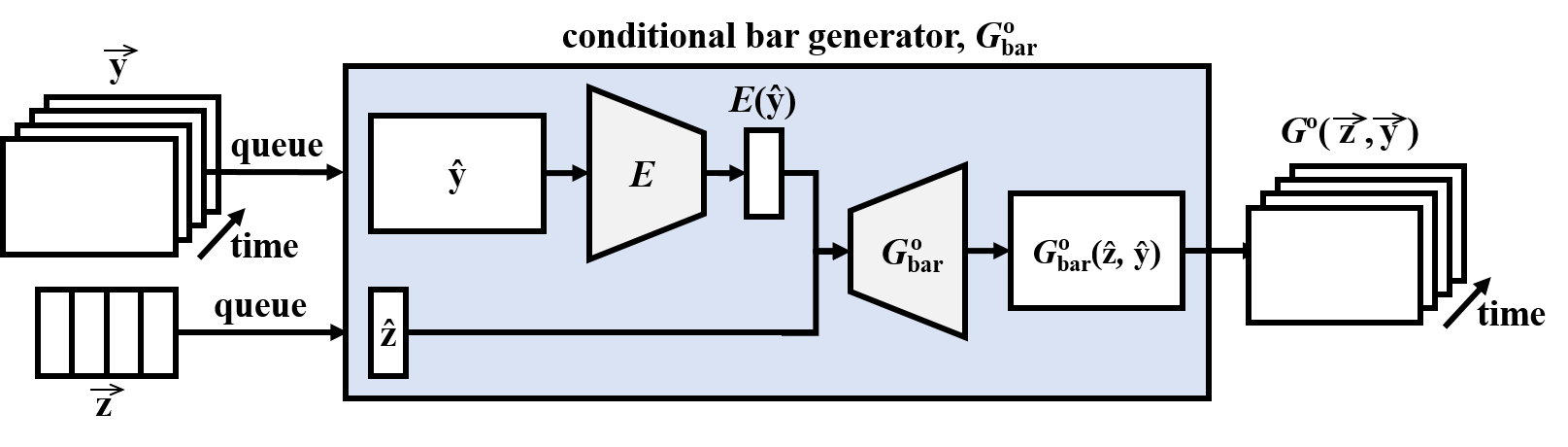} \\
(b) Track-conditional generation 
\caption{Two temporal models employed in our work. Note that only the generators are shown.}
\label{fig:temporal}
\end{figure}

\begin{figure*}[t]
\centering
\includegraphics[width=1.0\linewidth]{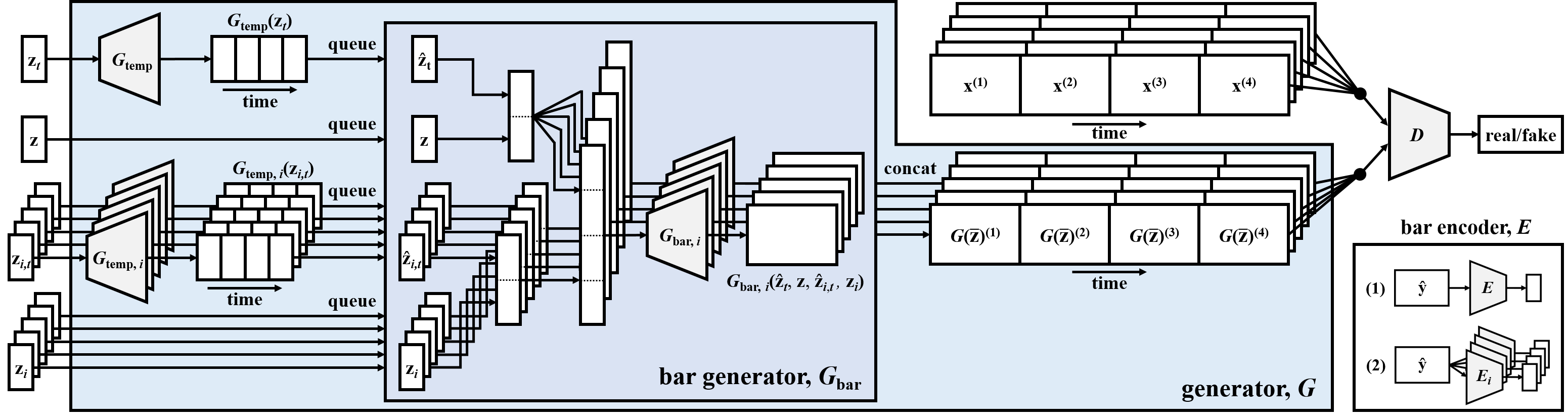}
\caption{System diagram of the proposed MuseGAN model for multi-track sequential data generation.}
\label{fig:musegan}
\end{figure*}

\subsection{Modeling the Temporal Structure}
\label{sec:temporal}

The models presented above can only generate multi-track music bar by bar, with possibly no coherence among the bars. We need a temporal model to generate music of a few bars long, such as a musical phrase (see Figure \ref{fig:hier}). We design two methods to achieve this, as described below.

\subsubsection{Generation from Scratch}
\label{sec:from-scratch}

The first method aims to generate fixed-length musical phrases by viewing bar progression as another dimension to grow the generator. The generator consists of two sub networks, the \emph{temporal structure generator} $G_{\t}$ and the \emph{bar generator} $G_{\b}$, as shown in Figure~\ref{fig:temporal}(a). $G_{\t}$ maps a noise vector $\z$ to a sequence of some latent vectors, $\overrightarrow{\z} = \{\overrightarrow{\z}^{(t)}\}_{t=1}^T$. The resulting $\overrightarrow{\z}$, which is expected to carry temporal information, is then be used by $G_{\b}$ to generate piano-rolls sequentially (i.e. bar by bar):
\begin{equation}
G\left(\z\right) = \left\{G_\b\left(G_\t\left(\z\right)^{(t)}\right)\right\}_{t=1}^T \,.
\end{equation}
We note that a similar idea has been used by \cite{TGAN} for video generation.

\subsubsection{Track-conditional Generation}
\label{sec:track-conditional}

The second method assumes that the bar sequence $\overrightarrow{\y}$ of one specific track is given by human, and tries to learn the temporal structure underlying that track and to generate the remaining tracks (and complete the song). As shown in Figure~\ref{fig:temporal}(b), the track-conditional generator $G^\circ$ generates bars one after another with the \emph{conditional bar generator}, $G^\circ_\text{bar}$. The multi-track piano-rolls of the remaining tracks of bar $t$ are then generated by $G^\circ_\text{bar}$, which takes two inputs, the condition $\overrightarrow{\y}^{(t)}$ and a \emph{time-dependent} random noise $\overrightarrow{\z}^{(t)}$.

In order to achieve such conditional generation with high-dimensional conditions, an additional encoder $E$ is trained to map $\overrightarrow{\y}^{(t)}$ to the space of $\overrightarrow{\z}^{(t)}$. Notably, similar approaches have been adopted by \cite{midinet}. The whole procedure can be formulated as
\begin{equation}
G^\circ\left(\overrightarrow{\z}, \overrightarrow{\y}\right) = \left\{G^\circ_\text{bar}\left(\overrightarrow{\z}^{(t)}, E \left(\overrightarrow{\y}^{(t)}\right)\right)\right\}_{t=1}^T \,.
\end{equation}

Note that the encoder is expected to extract inter-track features instead of intra-track features from the given track, since intra-track features are supposed not to be useful for generating the other tracks.

To our knowledge, incorporating a temporal model in this way is new. It can be applied to human-AI cooperative generation, or music accompaniment.

\subsection{MuseGAN}
\label{sec:musegan}

We now present the MuseGAN, an integration and extension of the proposed multi-track and temporal models. As shown in Figure \ref{fig:musegan}, the input to MuseGAN, denoted as $\mathbf{\bar{z}}$, is composed of four parts: an inter-track time-independent random vectors $\z$, an intra-track time-independent random vectors $\z_i$, an inter-track time-dependent random vectors $\z_t$ and an intra-track time-dependent random vectors $\z_{i,t}$.

For track $i$ ($i=1,2,\dots,M$), the shared temporal structure generator $G_\text{temp}$, and the private temporal structure generator $G_{\t,i}$ take the time-dependent random vectors, $\z_t$ and $\z_{i,t}$ respectively, as their inputs, and each of them outputs a series of latent vectors containing inter-track and intra-track, respectively, temporal information. The output series (of latent vectors), together with the time-independent random vectors, $\z$ and $\z_i$, are concatenated\footnote{Other vector operations such as summation are also feasible.} and fed to the bar generator $G_{\text{bar}}$, which then generates piano-rolls sequentially. The generation procedure can be formulated as 
\begin{equation}
\small
G(\bar{\z}) = \left\{G_{\b,i}\left(\z, G_{\t}(\z_t)^{(t)}, \z_i, G_{\t,i}(\z_{i,t})^{(t)} \right)\right\}_{i,t=1}^{M,T}\,.
\end{equation}

For the track-conditional scenario, an additional encoder $E$ is responsible for extracting useful inter-track features from the user-provided track.\footnote{One can otherwise use multiple encoders (see Figure \ref{fig:musegan}).} This can be done analogously so we omit the details due to space limitation.

\begin{figure*}[t]
\centering
\includegraphics[width=1.0\linewidth]{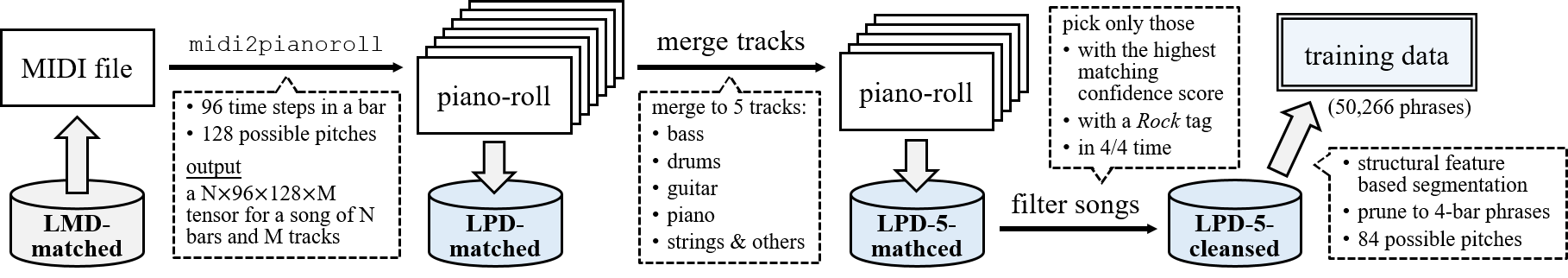}
\caption{Illustration of the dataset preparation and data preprocessing procedure.}
\label{fig:preprocessing}
\end{figure*}


\section{Implementation}
\label{sec:implementation}

\subsection{Dataset}
\label{sec:dataset}

The piano-roll dataset we use in this work is derived from the \emph{Lakh MIDI dataset} (LMD) \cite{raffel16phd},\footnote{\url{http://colinraffel.com/projects/lmd/}} a large collection of 176,581 unique MIDI files. We convert the MIDI files to multi-track piano-rolls. For each bar, we set the height to 128 and the width (time resolution) to 96 for modeling common temporal patterns such as triplets and 16th notes.\footnote{For tracks other than the drums, we enforce a rest of one time step at the end of each note to distinguish two successive notes of the same pitch from a single long note, and notes shorter than two time steps are dropped. For the drums, only the onsets are encoded.} We use the python library pretty\_midi \cite{pretty_midi} to parse and process the MIDI files. We name the resulting dataset the \emph{Lakh Pianoroll Dataset} (LPD). We also present the subset LPD-matched, which is derived from the LMD-matched, a subset of 45,129 MIDIs matched to entries in the Million Song Dataset (MSD) \cite{MSD}. Both datasets, along with the metadata and the conversion utilities, can be found on the project website.\footnotemark[1]

\subsection{Data Preprocessing}
\label{sec:preprocessing}

As these MIDI files are scraped from the web and mostly user-generated \cite{raffel16ismir}, the dataset is quite noisy. Hence, we use LPD-matched in this work and perform three steps for further cleansing (see Figure \ref{fig:preprocessing}).

First, some tracks tend to play only a few notes in the entire songs. This increases data sparsity and impedes the learning process. We deal with such a data imbalance issue by merging tracks of similar instruments (by summing their piano-rolls). Each multi-track piano-roll is compressed into five tracks: bass, drums, guitar, piano and strings.\footnote{Instruments out of the list are considered as part of the strings.} Doing so introduces noises to our data, but empirically we find it better than having empty bars. After this step, we get the \emph{LPD-5-matched}, which has 30,887 multi-track piano-rolls.

Since there is no clear way to identify which track plays the melody and which plays the accompaniment \cite{raffel16ismir}, we cannot categorize the tracks into melody, rhythm and drum tracks as some prior works did \cite{song_from_pi,midinet}.

Second, we utilize the metadata provided in the LMD and MSD, and we pick only the piano-rolls that have higher confidence score in matching,\footnote{The matching confidence comes with the LMD, which is the confidence of whether the MIDI file match any entry of the MSD.} that are \emph{Rock} songs and are in 4/4 time. After this step, we get the \emph{LPD-5-cleansed}.

Finally, in order to acquire musically meaningful phrases to train our temporal model, 
we segment the piano-rolls with a state-of-the-art algorithm, \emph{structural features} \cite{SF},\footnote{We use the MSAF toolbox \cite{MSAF} to run the algorithm: \url{https://github.com/urinieto/msaf}.} and obtain phrases accordingly. In this work, we consider four bars as a phrase and prune longer segments into proper size. We get 50,266 phrases in total for the training data. Notably, although we use our models to generate fixed-length segments only, the track-conditional model is able to generate music of any length according to the input.

Since very low and very high notes are uncommon, we discard notes below \texttt{C1} or above \texttt{C8}. The size of the target output tensor (i.e. the artificial piano-roll of a segment) is hence 4 (bar) $\times$ 96 (time step) $\times$ 84 (note) $\times$ 5 (track). (See Appendix~\ref{app:sec:sample_train} for sample piano-rolls in the training data.)

\begin{table*}[t]
\centering
\small
\begin{tabular}{|ll|ccccc|cccc|cccc|c|} 
  \hline
  & &\multicolumn{5}{c|}{empty bars (\textbf{EB}; \%)} &\multicolumn{4}{c|}{used pitch classes (\textbf{UPC})}  &\multicolumn{4}{c|}{qualified notes (\textbf{QN}; \%)} &\textbf{DP} (\%) 
  \\
  & &\textbf{B}&\textbf{D}&\textbf{G}&\textbf{P}&\textbf{S}
  &\textbf{B}&\textbf{G}&\textbf{P}&\textbf{S}
  &\textbf{B}&\textbf{G}&\textbf{P}&\textbf{S}
  &\textbf{D} \\ 
  \hline\hline
  \multicolumn{2}{|l|}{training data} &8.06 &8.06 &19.4 &24.8 &10.1 &1.71 &3.08 &3.28 &3.38 &90.0 &81.9 &88.4 &89.6 &88.6 \\ 
  \hline\hline
  \multirow{4}{*}{\shortstack[l]{from\\scratch}} 
  &jamming &6.59 &2.33 &18.3 &22.6 &6.10 &1.53 &3.69 &4.13 &4.09 &71.5 &56.6 &62.2 &63.1 &93.2 \\
  &composer &0.01 &28.9 &1.34 &0.02 &0.01 &2.51 &4.20 &4.89 &5.19 &49.5 &47.4 &49.9 &52.5 &75.3 \\
  &hybrid  &2.14 &29.7 &11.7 &17.8 &6.04 &2.35 &4.76 &5.45 &5.24 &44.6 &43.2 &45.5 &52.0 &71.3 \\
  \cdashline{2-16}[1pt/1pt]
  &ablated &92.4 &100 &12.5 &0.68 &0.00 &1.00 &2.88 &2.32 &4.72 &0.00 &22.8 &31.1 &26.2 &0.0  \\
  \hline
  \multirow{3}{*}{\shortstack[l]{track-\\conditional}}
  &jamming  &4.60 &3.47 &13.3 &--- &3.44 &2.05 &3.79 &--- &4.23 &73.9 &58.8 &--- &62.3 &91.6 \\
  &composer &0.65 &20.7 &1.97 &--- &1.49 &2.51 &4.57 &--- &5.10 &53.5 &48.4 &--- &59.0 &84.5 \\
  &hybrid   &2.09 &4.53 &10.3 &--- &4.05 &2.86 &4.43 &--- &4.32 &43.3 &55.6 &--- &67.1 &71.8 \\
  \hline
\end{tabular}
\caption{Intra-track evaluation (\textbf{B}: bass, \textbf{D}: drums, \textbf{G}: guitar, \textbf{P}: piano, \textbf{S}: strings; values closer to the first row are better)}
\label{tab:metrics_intra}
\end{table*}

\begin{table}[t]
\centering
\small
\begin{tabular}{|ll|cccccc|}
  \hline
  & &\multicolumn{6}{c|}{tonal distance (\textbf{TD})} \\
  &&B-G &B-S &B-P &G-S &G-P &S-P \\
  \hline\hline
  \multicolumn{2}{|l|}{train.} &1.57 &1.58 &1.51 &1.10 &1.02 &1.04 \\
  \multicolumn{2}{|l|}{train. (shuffled)} &1.59 &1.59 &1.56 &1.14 &1.12 &1.13 \\
  \hline\hline
  \multirow{3}{*}{\shortstack[l]{from\\scratch}} &jam. &1.56 &1.60 &1.54 &1.05 &0.99 &1.05 \\
  &comp. &1.37 &1.36 &\textbf{1.30} &0.95 &0.98 &0.91 \\
  &hybrid &\textbf{1.34} &\textbf{1.35} &1.32 &\textbf{0.85} &\textbf{0.85} &\textbf{0.83} \\
  \hline\hline
  \multirow{3}{*}{\shortstack[l]{track-\\condi-\\tional}} &jam.  &1.51 &1.53 &1.50 &1.04 &0.95 &1.00 \\
  &comp. &1.41 &\textbf{1.36} &1.40 &\textbf{0.96} &1.01 &\textbf{0.95} \\
  &hybrid  &\textbf{1.39} &\textbf{1.36} &\textbf{1.38} &\textbf{0.96} &\textbf{0.94} &\textbf{0.95} \\
  \hline
\end{tabular}
\caption{Inter-track evaluation (smaller values are better)}
\label{tab:metrics_inter}
\end{table}

\subsection{Model Settings}
\label{sec:model}

Both $G$ and $D$ are implemented as deep CNNs. $G$ grows the time axis first and then the pitch axis, while $D$ compresses in the opposite way. As suggested by \cite{ImprovedWGAN}, we update $G$ once every five updates of $D$ and apply batch normalization only to $G$. The total length of the input random vector(s) for each generator is fixed to 128.\footnote{It can be one single vector, two vectors of length 64 or four vectors of length 32, depending on the model employed.} The training time for each model is less than 24 hours with a Tesla K40m GPU. In testing stage, we binarize the output of $G$, which uses tanh as activation functions in the last layer, by a threshold at zero. (See Appendix \ref{app:sec:implementation} for more details.)


\section{Objective Metrics for Evaluation}
\label{sec:metrics}

To evaluate our models, we design several metrics that can be computed for both the real and the generated data, including four intra-track and one inter-track (the last one) metrics:
\begin{itemize}
\item \textbf{EB}: ratio of empty bars (in \%).
\item \textbf{UPC}: number of used pitch classes per bar (from 0 to 12).
\item \textbf{QN}: ratio of ``qualified'' notes  (in \%). We consider a note no shorter than three time steps (i.e. a 32th note) as a qualified note. QN shows if the music is overly fragmented.
\item \textbf{DP}, or drum pattern: ratio of notes in 8- or 16-beat patterns, common ones for Rock songs in 4/4 time (in \%).
\item \textbf{TD}: or tonal distance \cite{tonalDist}. It measures the hamornicity between a pair of tracks. Larger TD implies weaker inter-track harmonic relations.
\end{itemize}

By comparing the values computed from the real and the fake data, we can get an idea of the performance of generators. The concept is similar to the one in GANs---the distributions (and thus the statistics) of the real and the fake data should become closer as the training process proceeds.

\subsection{Analysis of Training Data}

We apply these metrics to the training data to gain a greater understanding of our training data. The result is shown in the first rows of Tables \ref{tab:metrics_intra} and \ref{tab:metrics_inter}. The values of \textbf{EB} show that categorizing the tracks into five families is appropriate. From \textbf{UPC}, we find that the bass tends to play the melody, which results in a UPC below 2.0, while the guitar, piano and strings tend to play the chords, which results in a UPC above 3.0. High values of \textbf{QN} indicate that the converted piano-rolls are not overly fragmented. From \textbf{DP}, we see that over 88 percent of the drum notes are in either 8- or 16-beat patterns. The values of \textbf{TD} are around 1.50 when measuring the distance between a melody-like track (mostly the bass) and a chord-like track (mostly one of the piano, guitar or strings), and around 1.00 for two chord-like tracks. Notably, TD will slightly increase if we shuffle the training data by randomly pairing bars of two specific tracks, which shows that TD are indeed capturing inter-track harmonic relations.


\section{Experiment and Results}
\label{sec:exp}

\subsection{Example Results}
\label{sec:examples}

Figure \ref{fig:sample_fs} shows the piano-rolls of six phrases generated by the composer and the hybrid model. (See Appendix \ref{app:sec:samples} for more piano-roll samples.) Some rendered audio samples can be found on our project website.\footnotemark[1]

Some observations:
\begin{itemize}
\item The tracks are usually playing in the same music scale.
\item Chord-like intervals can be observed in some samples.
\item The bass often plays the lowest pitches and it is monophonic at most time (i.e. playing the melody).
\item The drums usually have 8- or 16-beat rhythmic patterns.
\item The guitar, piano and strings tend to play the chords, and their pitches sometimes overlap (creating the black lines), which indicates nice harmonic relations.
\end{itemize}

\subsection{Objective Evaluation}
\label{sec:objective}

To examine our models, we generate 20,000 bars with each model and evaluate them in terms of the proposed objective metrics. The result is shown in Tables \ref{tab:metrics_intra} and \ref{tab:metrics_inter}. Note that for the conditional generation scenario, we use the piano tracks as conditions and generate the other four tracks. For comparison, we also include the result of an \emph{ablated} version of the composer model, one without batch normalization layers. This ablated model barely learns anything, so its values can be taken as a reference.

For the intra-track metrics, we see that the jamming model tends to perform the best. This is possibly because each generator in the jamming model is designed to focus on its own track only. Except for the ablated one, all models perform well in \textbf{DP}, which suggests that the drums do capture some rhythmic patterns in the training data, despite the relatively high \textbf{EB} for drums in the composer and the hybrid model. From \textbf{UPC} and \textbf{QN}, we see that all models tend to use more pitch classes and produce fairly less qualified notes than the training data do. This indicates that some noise might have been produced and that the generated music contains a great amount of overly fragmented notes, which may result from the way we binarize the continuous-valued output of $G$ (to create binary-valued piano-rolls). We do not have a smart solution yet and leave this as a future work.

\begin{figure}[t]
\centering
\includegraphics[width=1.0\linewidth, height=4cm]{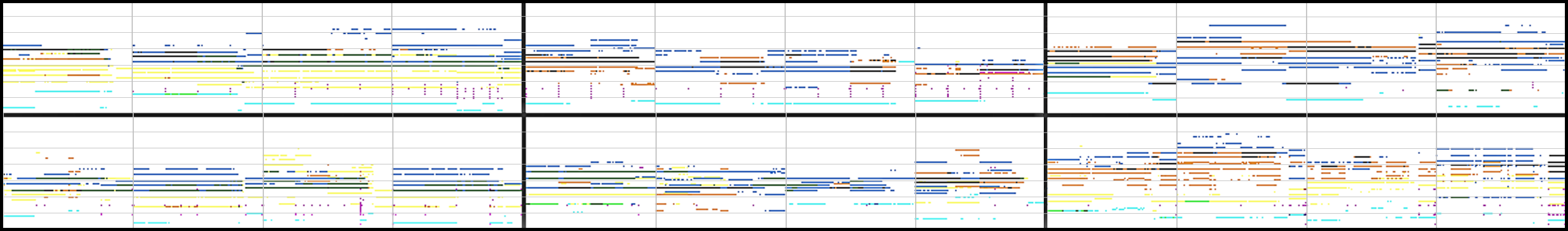}
\caption{Example generative results for the composer model (top row) and the hybrid model (bottom row), both generating from scratch (best viewed in color---\textbf{cyan}: bass, \textbf{pink}: drums, \textbf{yellow}: guitar, \textbf{blue}: strings, \textbf{orange}: piano)}
\label{fig:sample_fs}
\end{figure}

For the inter-track metric \textbf{TD} (Table \ref{tab:metrics_inter}), we see that the values for the composer model and the hybrid model are relatively lower than that of the jamming models. This suggests that the music generated by the jamming model has weaker harmonic relation among tracks and that the composer model and the hybrid model may be more appropriate for multi-track generation in terms of cross-track harmonic relation. Moreover, we see that composer model and the hybrid model perform similarly across different combinations of tracks. This is encouraging for we know that the hybrid model may not have traded performance for its flexibility.

\begin{figure*}[t]
\centering
\begin{subfigure}[]{0.05\textwidth}
  \centering
  bass \\ [0.33cm]
  drums \\ [0.33cm]
  guitar \\ [0.33cm]
  strings \\ [0.33cm]
  piano \\ \vspace{0.5cm}
\end{subfigure}
\hfill
\begin{subfigure}[]{0.184\textwidth}
  \centering
  \includegraphics[scale=0.315]{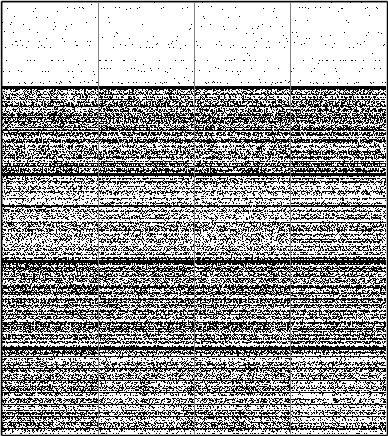}
  step 0 (A)
\end{subfigure}
\begin{subfigure}[]{0.184\textwidth}
  \centering
  \includegraphics[scale=0.315]{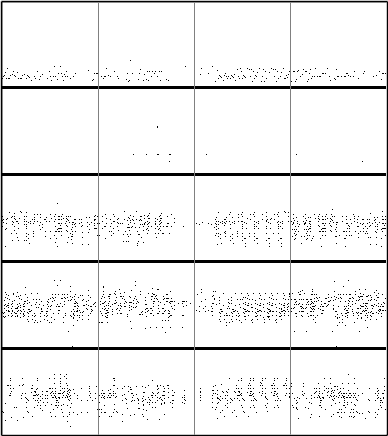}
  step 700 (B)
\end{subfigure}
\begin{subfigure}[]{0.184\textwidth}
  \centering
  \includegraphics[scale=0.315]{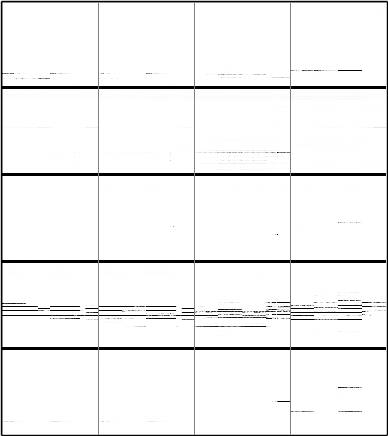}
  step 2500 (C)
\end{subfigure}
\begin{subfigure}[]{0.184\textwidth}
  \centering
  \includegraphics[scale=0.315]{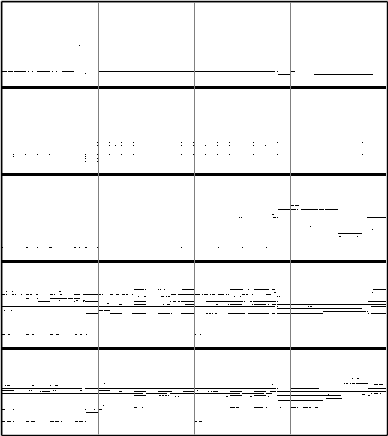}
  step 6000 (D)
\end{subfigure}
\begin{subfigure}[]{0.184\textwidth}
  \centering
  \includegraphics[scale=0.315]{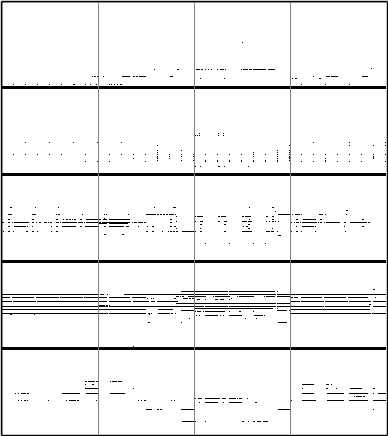}
  step 7900 (E)
\end{subfigure}
\caption{Evolution of the generated piano-rolls as a function of update steps, for the composer model generating from scratch.}
\label{fig:evolution}
\end{figure*}

\begin{figure}[t]
\centering
(a)\includegraphics[width=.95\linewidth]{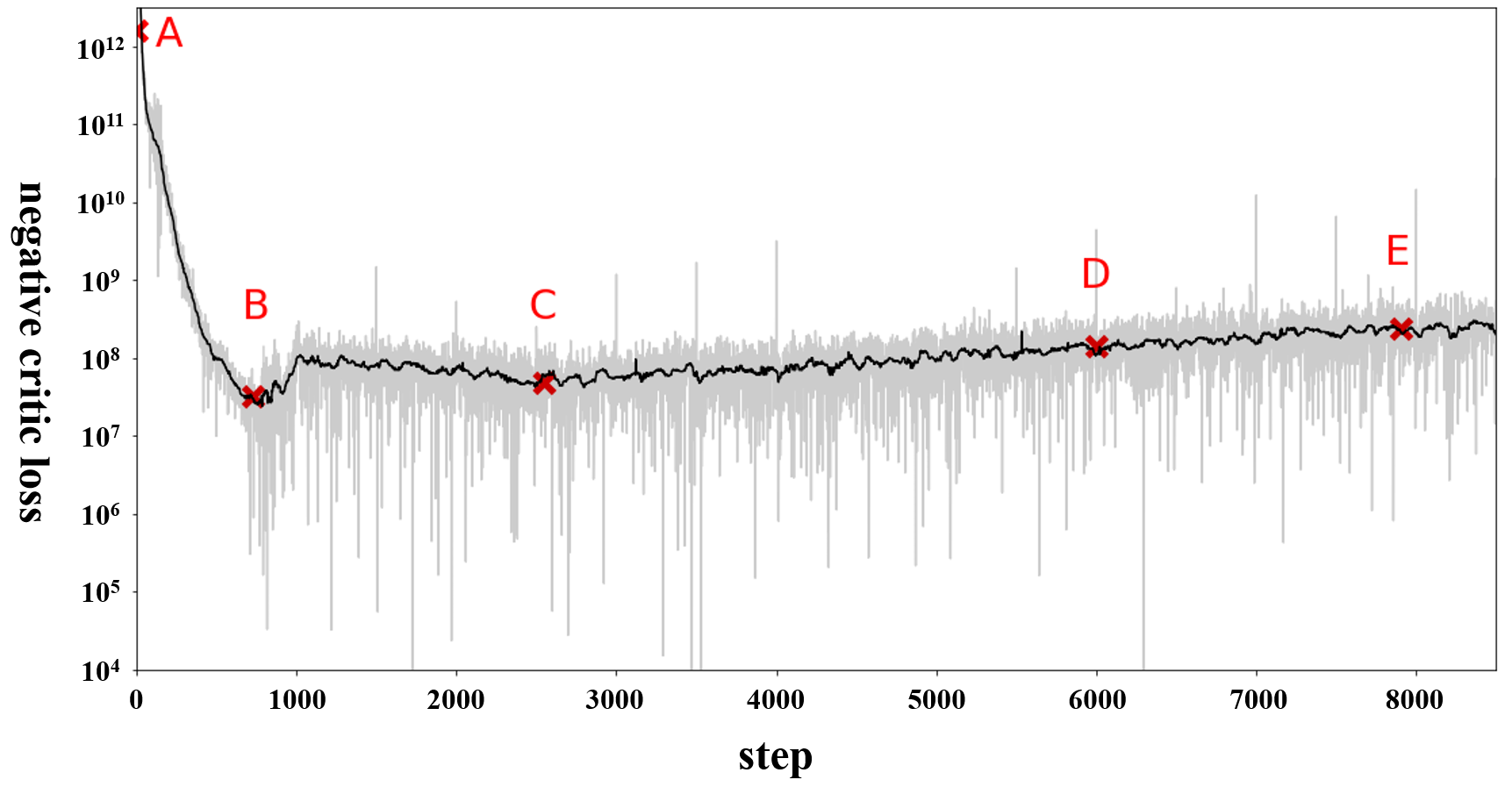} \\
\vspace{0.2cm}
\begin{subfigure}[]{0.49\linewidth}
  \centering
  (b)\includegraphics[width=0.9\linewidth]{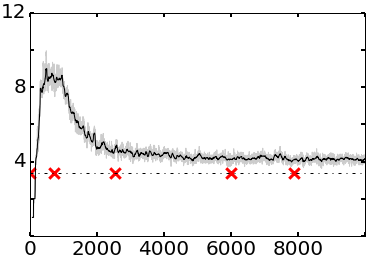} \\
\end{subfigure}
\hfill
\begin{subfigure}[]{0.49\linewidth}
  \centering
  (c)\includegraphics[width=0.9\linewidth]{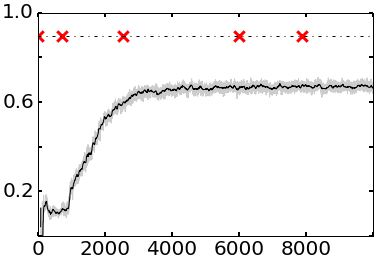} \\
\end{subfigure}
\caption{(a) Training loss of the discriminator, (b) the \textbf{UPC} and (c) the \textbf{QN} of the strings track, for the composer model generating from scratch. The gray and black curves are the raw values and the smoothed ones (by median filters), respectively. The dashed lines in (b) and (c) indicate the values calculated from the training data.}
\label{fig:train_variation}
\end{figure}

\subsection{Training Process}
\label{train_process}

To gain insights of the training process, we firstly study the composer model for generation from scratch (other models have similar behaviors). Figure \ref{fig:train_variation}(a) shows the training loss of $D$ as a function of training steps. We see that it decreases rapidly in the beginning and then saturates. However, there is a mild growing trend after point B marked on the graph, suggesting that $G$ starts to learn something after that.

We show in Figure \ref{fig:evolution} the generated piano-rolls at the five points marked on Figure \ref{fig:train_variation}(a), from which we can observe how the generated piano-rolls evolve as the training process unfolds. For example, we see that $G$ grasps the pitch range of each track quite early and starts to produce some notes, fragmented but within proper pitch ranges, at point B rather than noises produced at point A. At point B, we can already see cluster of points gathering at the lower part (with lower pitches) of the bass. After point C, we see that the guitar, piano and strings start to learn the duration of notes and begin producing longer notes. These results show that $G$ indeed becomes better as the training process proceeds.

We also show in Figure \ref{fig:train_variation} the values of two objective metrics along the training process. From (b) we see that $G$ can ultimately learn the proper number of pitch classes; from (c) we see that QN stays fairly lower than that of the training data, which suggests room for further improving our $G$. These show that a researcher can employ these metrics to study the generated result, before launching a subjective test.

\begin{table}[t]
\centering
\small
\begin{tabular}{|l|ll|ccccc|}
  \hline
  \multicolumn{3}{|c|}{} &\textbf{H} &\textbf{R} &\textbf{MS} &\textbf{C} &\textbf{OR} \\
  \hline\hline
  \multirow{6}{*}{\rotatebox{90}{\shortstack{from\\scratch}}}
  &\multirow{3}{*}{\shortstack[l]{non-\\pro}}
    &jam.   &2.83 &3.29 &2.88 &2.84 &2.88 \\
  & &comp.  &3.12 &\textbf{3.36} &2.95 &3.13 &3.12 \\
  & &hybrid &\textbf{3.15} &3.33 &\textbf{3.09} &\textbf{3.30} &\textbf{3.16} \\
  \cline{2-8}
  &\multirow{3}{*}{pro}
    &jam.   &2.31 &3.05 &2.48 &2.49 &2.42 \\
  & &comp.  &2.66 &3.13 &2.68 &2.63 &2.73 \\
  & &hybrid &\textbf{2.92} &\textbf{3.25} &\textbf{2.81} &\textbf{3.00} &\textbf{2.93} \\
  \hline\hline
  \multirow{6}{*}{\rotatebox{90}{\shortstack{track-\\conditional}}}
  &\multirow{3}{*}{\shortstack[l]{non-\\pro}}
    &jam.   &\textbf{2.89} &\textbf{3.44} &2.97 &\textbf{3.01} &\textbf{3.06} \\
  & &comp.  &2.70 &3.29 &\textbf{2.98} &2.97 &2.86 \\
  & &hybrid &2.78 &3.34 &2.93 &2.98 &3.01 \\
  \cline{2-8}
  &\multirow{3}{*}{pro}
    &jam.   &2.44 &\textbf{3.32} &2.67 &2.72 &2.69 \\
  & &comp.  &2.35 &3.21 &2.59 &2.67 &2.62 \\
  & &hybrid &\textbf{2.49} &3.29 &\textbf{2.71} &\textbf{2.73} &\textbf{2.70} \\
  \hline
\end{tabular}
\caption{Result of user study (\textbf{H}: harmonious, \textbf{R}: rhythmic, \textbf{MS}: musically structured, \textbf{C}: coherent, \textbf{OR}: overall rating)}
\label{tab:user_study}
\end{table}


\subsection{User Study}
\label{sec:user_study}

Finally, we conduct a listening test of 144 subjects recruited from the Internet via our social circles. 44 of them are deemed `pro user,' according to a simple questionnaire probing their musical background. Each subject has to listen to nine music clips in random order. Each clip consists of three four-bar phrases generated by one of the proposed models and quantized by sixteenth notes. The subject rates the clips in terms of whether they 1) have pleasant harmony, 2) have unified rhythm, 3) have clear musical structure, 4) are coherent, and 5) the overall rating, in a 5-point Likert scale.

From the result shown in Table~\ref{tab:user_study}, the hybrid model is preferred by pros and non-pros for generation from scratch and by pros for conditional generation, while the jamming model is preferred by non-pros for conditional generation. Moreover, the composer and the hybrid models receive higher scores for the criterion Harmonious for generation from scratch than the jamming model does, which suggests that the composer and the hybrid models perform better at handling inter-track interdependency.


\section{Related Work}
\label{section:related_work}

\subsection{Video Generation using GANs}
\label{sec:video_gen}

Similar to music generation, a temporal model is also needed for video generation. Our model design is inspired by some prior arts that used GANs in video generation. VGAN \cite{VGAN} assumed that a video can be decomposed into a dynamic foreground and a static background. They used 3D and 2D CNNs to generate them respectively in a two-stream architecture and combined the results via a mask generated by the foreground stream. TGAN \cite{TGAN} used a temporal generator (using convolutions) to generate a fixed-length series of latent variables, which is then be fed one by one to an image generator to generate the video frame by frame. MoCoGAN \cite{MoCoGAN} assumed that a video can be decomposed into content (objects) and motion (of objects) and used RNNs to capture the motion of objects.

\subsection{Symbolic Music Generation}
\label{sec:music_gen}

As reviewed by \cite{briot17survey}, a surging number of models have been proposed lately for symbolic music generation. Many of them used RNNs to generate music of different formats, including monophonic melodies \cite{folkrnn} and four-voice chorales \cite{DBach}. Notably, RNN-RBM \cite{RBM}, a generalization of the recurrent temporal restricted Boltzmann machine (RTRBM), was able to generate polyphonic piano-rolls of a single track. Song from PI \cite{song_from_pi} were able to generate a lead sheet (i.e. a track of melody and a track of chord tags) with an additional monophonic drums track by using hierarchical RNNs to coordinate the three tracks.

Some recent works have also started to explore using GANs for generating music. C-RNN-GAN \cite{crnn} generated polyphonic music as a series of note events\footnote{In the \emph{note event representation}, music is viewed as a series of note event, which is typically denoted as a tuple of onset time, pitch, velocity and duration (or offset time).} by introducing some ordering of notes and using RNNs in both the generator and the discriminator. SeqGAN \cite{SeqGAN} combined GANs and reinforcement learning to generate sequences of discrete tokens. It has been applied to generate monophonic music, using the note event representation.\footnotemark[10] MidiNet \cite{midinet} used conditional, convolutional GANs to generate melodies that follows a chord sequence given a priori, either from scratch or conditioned on the melody of previous bars.


\section{Conclusion}
\label{sec:conclusion}

In this work, we have presented a novel generative model for multi-track sequence generation under the framework of GANs. We have also implemented such a model with deep CNNs for generating multi-track piano-rolls. We designed several objective metrics and showed that we can gain insights into the learning process via these objective metrics. The objective metrics and the subjective user study show that the proposed models can start to learn something about music. Although musically and aesthetically it may still fall behind the level of human musicians, the proposed model has a few desirable properties, and we hope follow-up research can further improve it.

\bibliography{ref.bib}

\begin{thebibliography}{}

\bibitem[\protect\citeauthoryear{Arjovsky, Chintala, and Bottou}{2017}]{WGAN}
Arjovsky, M.; Chintala, S.; and Bottou, L.
\newblock 2017.
\newblock Wasserstein {GAN}.
\newblock {\em arXiv preprint arXiv:1701.07875}.

\bibitem[\protect\citeauthoryear{Bertin-Mahieux \bgroup et al\mbox.\egroup
  }{2011}]{MSD}
Bertin-Mahieux, T.; Ellis, D.~P.; Whitman, B.; and Lamere, P.
\newblock 2011.
\newblock The {M}illion {S}ong {D}ataset.
\newblock In {\em ISMIR}.

\bibitem[\protect\citeauthoryear{Boulanger-Lewandowski, Bengio, and
  Vincent}{2012}]{RBM}
Boulanger-Lewandowski, N.; Bengio, Y.; and Vincent, P.
\newblock 2012.
\newblock Modeling temporal dependencies in high-dimensional sequences:
  Application to polyphonic music generation and transcription.
\newblock In {\em ICML}.

\bibitem[\protect\citeauthoryear{Briot, Hadjeres, and
  Pachet}{2017}]{briot17survey}
Briot, J.-P.; Hadjeres, G.; and Pachet, F.
\newblock 2017.
\newblock Deep learning techniques for music generation: A survey.
\newblock {\em arXiv preprint arXiv:1709.01620}.

\bibitem[\protect\citeauthoryear{Chu, Urtasun, and Fidler}{2017}]{song_from_pi}
Chu, H.; Urtasun, R.; and Fidler, S.
\newblock 2017.
\newblock Song from {PI}: {A} musically plausible network for pop music
  generation.
\newblock In {\em ICLR Workshop}.

\bibitem[\protect\citeauthoryear{Goodfellow \bgroup et al\mbox.\egroup
  }{2014}]{GAN}
Goodfellow, I.~J.; Pouget-Abadie, J.; Mirza, M.; Xu, B.; Warde-Farley, D.;
  Ozair, S.; Courville, A.; and Bengio, Y.
\newblock 2014.
\newblock Generative adversarial nets.
\newblock In {\em NIPS}.

\bibitem[\protect\citeauthoryear{Gulrajani \bgroup et al\mbox.\egroup
  }{2017}]{ImprovedWGAN}
Gulrajani, I.; Ahmed, F.; Arjovsky, M.; Dumoulin, V.; and Courville, A.
\newblock 2017.
\newblock Improved training of {W}asserstein {GAN}s.
\newblock {\em arXiv preprint arXiv:1704.00028}.

\bibitem[\protect\citeauthoryear{Hadjeres, Pachet, and Nielsen}{2017}]{DBach}
Hadjeres, G.; Pachet, F.; and Nielsen, F.
\newblock 2017.
\newblock {DeepBach}: A steerable model for {B}ach chorales generation.
\newblock In {\em ICML}.

\bibitem[\protect\citeauthoryear{Harte, Sandler, and Gasser}{2006}]{tonalDist}
Harte, C.; Sandler, M.; and Gasser, M.
\newblock 2006.
\newblock Detecting harmonic change in musical audio.
\newblock In {\em ACM MM workshop on Audio and music computing multimedia}.

\bibitem[\protect\citeauthoryear{Herremans and Chew}{2017}]{herremans17tac}
Herremans, D., and Chew, E.
\newblock 2017.
\newblock {MorpheuS}: generating structured music with constrained patterns and
  tension.
\newblock {\em IEEE Trans. Affective Computing}.

\bibitem[\protect\citeauthoryear{Mogren}{2016}]{crnn}
Mogren, O.
\newblock 2016.
\newblock {C-RNN-GAN}: Continuous recurrent neural networks with adversarial
  training.
\newblock In {\em NIPS Worshop on Constructive Machine Learning Workshop}.

\bibitem[\protect\citeauthoryear{Nieto and Bello}{2016}]{MSAF}
Nieto, O., and Bello, J.~P.
\newblock 2016.
\newblock Systematic exploration of computational music structure research.
\newblock In {\em ISMIR}.

\bibitem[\protect\citeauthoryear{Radford, Metz, and Chintala}{2016}]{DCGAN}
Radford, A.; Metz, L.; and Chintala, S.
\newblock 2016.
\newblock Unsupervised representation learning with deep convolutional
  generative adversarial networks.
\newblock In {\em ICLR}.

\bibitem[\protect\citeauthoryear{Raffel and Ellis}{2014}]{pretty_midi}
Raffel, C., and Ellis, D. P.~W.
\newblock 2014.
\newblock Intuitive analysis, creation and manipulation of {MIDI} data with
  pretty\_midi.
\newblock In {\em ISMIR Late Breaking and Demo Papers}.

\bibitem[\protect\citeauthoryear{Raffel and Ellis}{2016}]{raffel16ismir}
Raffel, C., and Ellis, D. P.~W.
\newblock 2016.
\newblock Extracting ground truth information from {MIDI} files: A {MIDI}festo.
\newblock In {\em ISMIR}.

\bibitem[\protect\citeauthoryear{Raffel}{2016}]{raffel16phd}
Raffel, C.
\newblock 2016.
\newblock {\em Learning-Based Methods for Comparing Sequences, with
  Applications to Audio-to-MIDI Alignment and Matching}.
\newblock Ph.D. Dissertation, Columbia University.

\bibitem[\protect\citeauthoryear{Saito, Matsumoto, and Saito}{2017}]{TGAN}
Saito, M.; Matsumoto, E.; and Saito, S.
\newblock 2017.
\newblock Temporal generative adversarial nets with singular value clipping.
\newblock In {\em ICCV}.

\bibitem[\protect\citeauthoryear{Serr\`{a} \bgroup et al\mbox.\egroup
  }{2012}]{SF}
Serr\`{a}, J.; Müller, M.; Grosche, P.; and Arcos, J.~L.
\newblock 2012.
\newblock Unsupervised detection of music boundaries by time series structure
  features.
\newblock In {\em AAAI}.

\bibitem[\protect\citeauthoryear{Sturm \bgroup et al\mbox.\egroup
  }{2016}]{folkrnn}
Sturm, B.~L.; Santos, J.~F.; Ben{-}Tal, O.; and Korshunova, I.
\newblock 2016.
\newblock Music transcription modelling and composition using deep learning.
\newblock In {\em Conference on Computer Simulation of Musical Creativity}.

\bibitem[\protect\citeauthoryear{Tulyakov \bgroup et al\mbox.\egroup
  }{2017}]{MoCoGAN}
Tulyakov, S.; Liu, M.; Yang, X.; and Kautz, J.
\newblock 2017.
\newblock {MoCoGAN}: Decomposing motion and content for video generation.
\newblock {\em arXiv preprint arXiv:1707.04993}.

\bibitem[\protect\citeauthoryear{Vondrick, Pirsiavash, and
  Torralba}{2016}]{VGAN}
Vondrick, C.; Pirsiavash, H.; and Torralba, A.
\newblock 2016.
\newblock Generating videos with scene dynamics.
\newblock In {\em NIPS}.

\bibitem[\protect\citeauthoryear{Yang, Chou, and Yang}{2017}]{midinet}
Yang, L.-C.; Chou, S.-Y.; and Yang, Y.-H.
\newblock 2017.
\newblock {MidiNet}: A convolutional generative adversarial network for
  symbolic-domain music generation.
\newblock In {\em ISMIR}.

\bibitem[\protect\citeauthoryear{Yu \bgroup et al\mbox.\egroup }{2017}]{SeqGAN}
Yu, L.; Zhang, W.; Wang, J.; and Yu, Y.
\newblock 2017.
\newblock Seq{GAN}: Sequence generative adversarial nets with policy gradient.
\newblock In {\em AAAI}.

\end{thebibliography}
\bibliographystyle{aaai}


\newpage

\begin{appendices}
\setcounter{secnumdepth}{1}

\section{Samples of the Training Data}
\label{app:sec:sample_train}

\begin{figure*}[t]
\centering
\includegraphics[width=\linewidth]{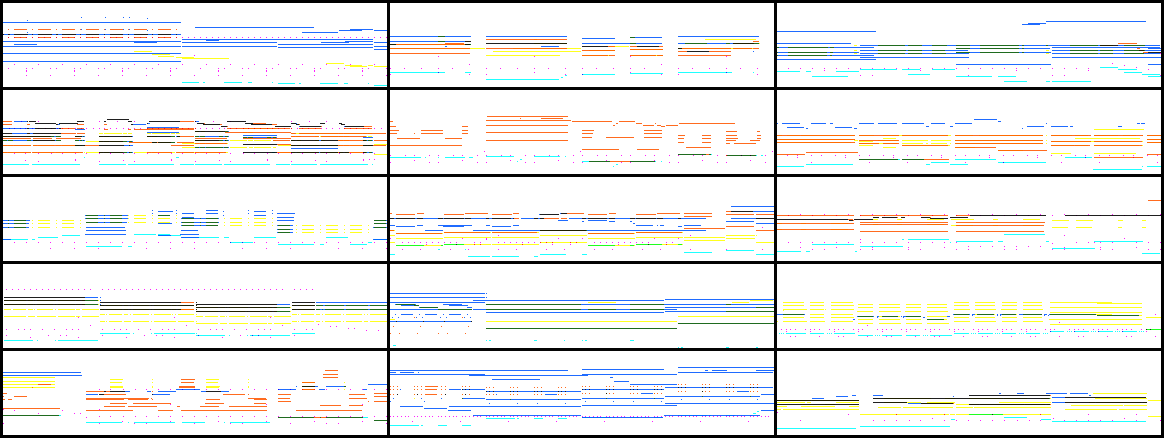}
\caption{Sample piano-rolls in the training data (best viewed in color---\textbf{cyan}: bass, \textbf{pink}: drums, \textbf{yellow}: guitar, \textbf{blue}: strings, \textbf{orange}: piano)}
\label{fig:train_more}
\end{figure*}

We show in Figure~\ref{fig:train_more} some randomly-chosen sample piano-rolls in the training data.

\section{Implementation Details}
\label{app:sec:implementation}

The network architecture of the proposed model is tabulated in Table \ref{tab:arch}. Some details can be found below.

\subsection{Random Vectors}
The total length of the input random vector(s) for the whole system is fixed to 128, which can be one single vector, two vectors of length 64 or four vectors of length 32, depending on the model employed. The input random vector of $G_\t$ has the same length as its output latent vectors. Thus, the total length of the input vector(s) of $G_\b$ is 128 as well.

\subsection{Network Architectures}

\subsubsection{Generator}
$G_\t$ consists of two 1-D transposed convolutional layers along the (inter-bar) time axis. $G_\b$ is composed of five 1-D transposed convolutional layers along the (intra-bar) time axis and two 1-D transposed convolutional layers along the pitch axis successively. A batch normalization (BN) layer is added before each activation layer.

\subsubsection{Discriminator}
$D$ consists of five 1-D convolutional layers and one fully-connected layer. The negative slope of the leaky rectified linear units (ReLU) is set to 0.2.

\subsubsection{Encoder} $E$ has a reverse architecture as $G$, and \emph{skip connections} are applied to the corresponding layers in order to speed up the training process. We constrain the number of filters to 16 for each layer to compress the representation of inter-track interdependency.

\subsection{Training}
We train the whole network end-to-end using the Adam optimizer with $\alpha=0.001$, $\beta_1=0.5$, $\beta_2=0.9$. As suggested by \cite{ImprovedWGAN}, we update $G$ (and $E$ for the track-conditional generation model) once every five updates of $D$. The training time for each model is less than 24 hours with a Tesla K40m GPU.

\subsection{Rendering Audios}
First, we quantize the generated piano-rolls by sixteenth notes to avoid overly fragmented notes. After that, we convert the piano-rolls into MIDI files. The tracks are then mixed and rendered to stereo audio files in an external digital audio workstation.

\section{Sample Generated Piano-rolls}
\label{app:sec:samples}

\begin{figure*}[t]
\centering
\includegraphics[width=\linewidth]{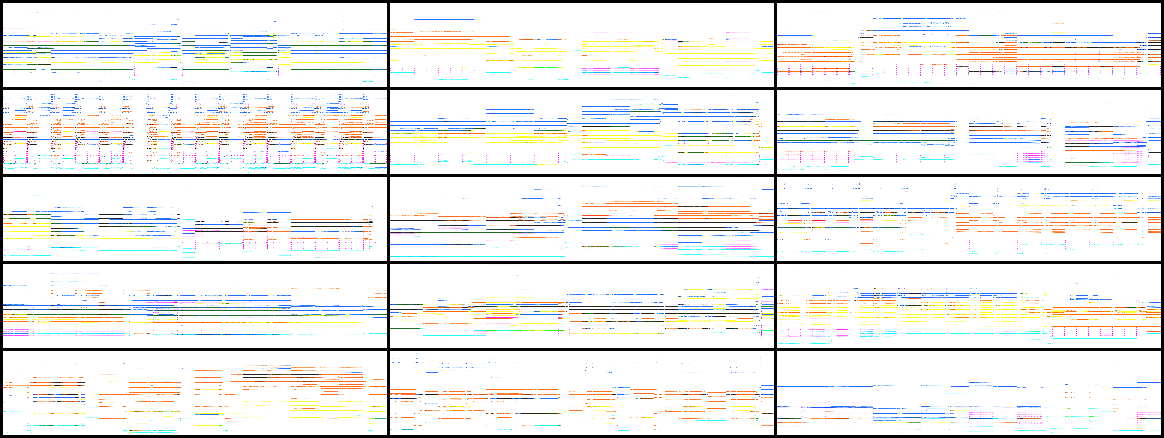} \\
(a) original piano-rolls (before binarization) \\
\vspace{0.5cm}
\includegraphics[width=\linewidth]{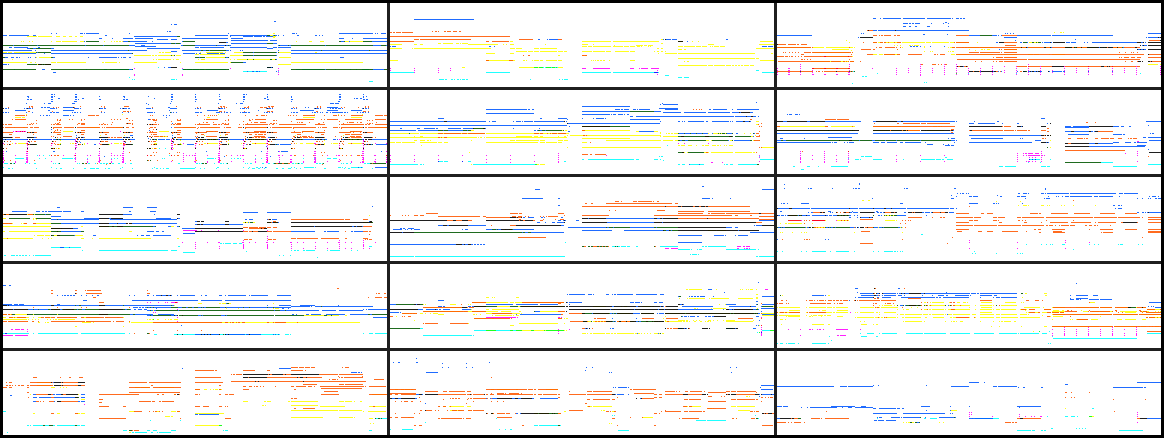} \\
(b) binarized piano-rolls
\caption{Randomly-chosen piano-rolls generated from scratch by the composer model (best viewed in color---\textbf{cyan}: bass, \textbf{pink}: drums, \textbf{yellow}: guitar, \textbf{blue}: strings, \textbf{orange}: piano). In (b) we binarize the output of $G$, which uses tanh as activation functions in the last layer, by a threshold at zero.}
\label{fig:comp_more}
\end{figure*}

\begin{figure*}[t]
\centering
\includegraphics[width=\linewidth]{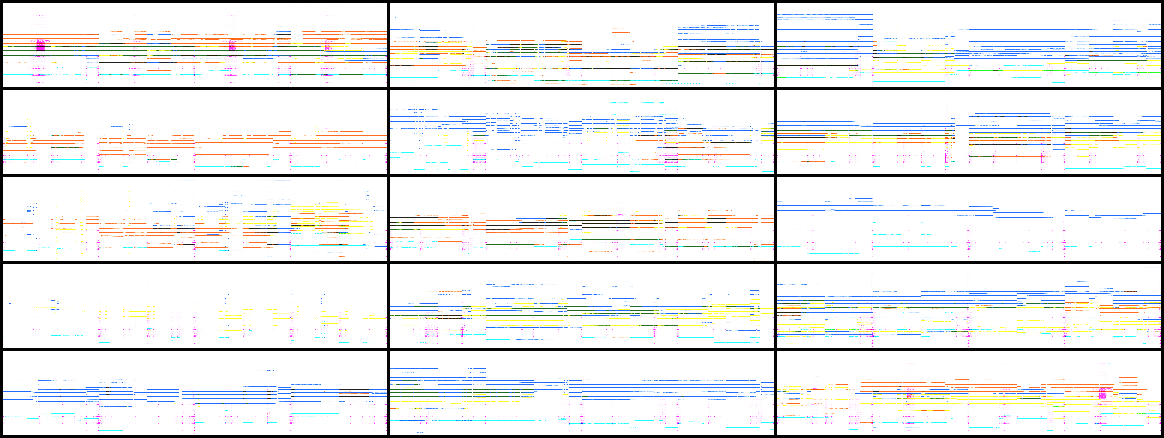} \\
(a) original piano-rolls (before binarization) \\
\vspace{0.5cm}
\includegraphics[width=\linewidth]{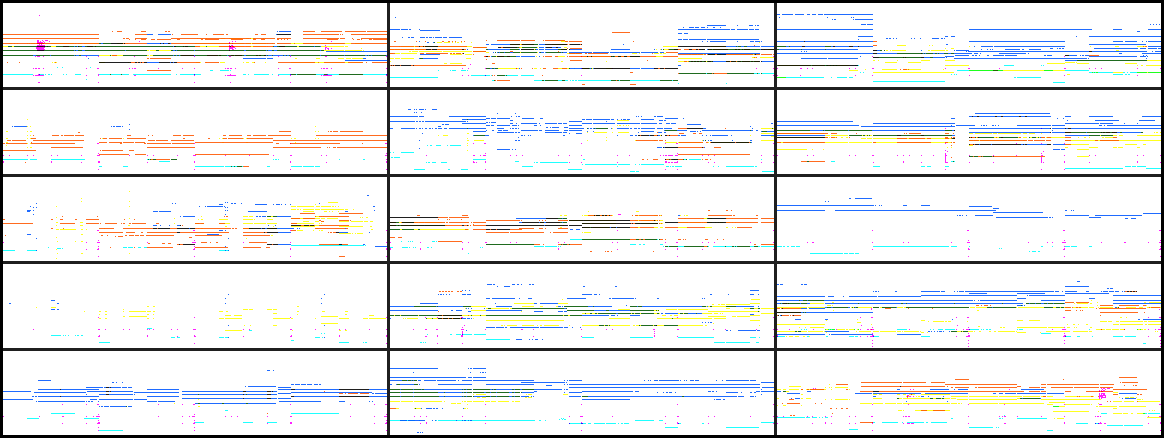} \\
(b) binarized piano-rolls
\caption{Randomly-chosen piano-rolls generated from scratch by the hybrid model (best viewed in color---\textbf{cyan}: bass, \textbf{pink}: drums, \textbf{yellow}: guitar, \textbf{blue}: strings, \textbf{orange}: piano)}
\label{fig:hybrid_more}
\end{figure*}

\begin{figure*}[t]
\centering
\includegraphics[width=\linewidth]{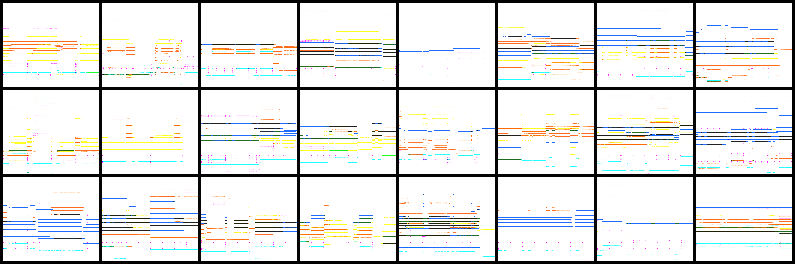} \\
(a) original piano-rolls (before binarization) \\
\vspace{0.5cm}
\includegraphics[width=\linewidth]{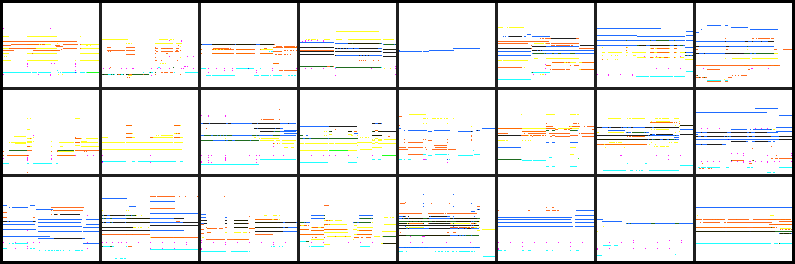} \\
(b) binarized piano-rolls
\caption{Randomly-chosen generated piano-rolls for the composer model conditioned on the strings track (best viewed in color---\textbf{cyan}: bass, \textbf{pink}: drums, \textbf{yellow}: guitar, \textbf{blue}: strings (conditions), \textbf{orange}: piano)}
\label{fig:cond_more}
\end{figure*}

We provide examples of randomly-chosen piano-rolls generated by our models.
\begin{itemize}
\item Figures~\ref{fig:comp_more} and \ref{fig:hybrid_more} show samples generated from scratch by the composer and the hybrid models, respectively.
\item Figure~\ref{fig:cond_more} shows samples of track-conditional generation for composer model. Note that we use the strings track, instead of the piano track used in the Experiment section, as conditions here to show the flexibility of our models.
\end{itemize}

\begin{table}[t]
\centering
\begin{tabular}{|l c c c c r|}
  \hline
  \multicolumn{6}{|l|}{Input: $\z \in \Re^{32}$} \\
  \hline
  \hline
  \multicolumn{6}{|l|}{reshaped to (1)$\times$32 channels} \\
  \hdashline[1pt/1pt]
  transconv &~~~~1024~~~~ &~~~2~~~ &~~~2~~~ &~BN~ &ReLU \\
  transconv &$K_\t$ &3 &1 &BN &ReLU \\
  \hline
  \hline
  \multicolumn{6}{|l|}{Output: $G_\t\left(\z\right) \in \Re^{32\times K_\t}$ ($K_\t$-track latent vector)} \\
  \hline
\end{tabular} \\
\vspace{0.15cm}
(a) the temporal generator $G_\t$
\vspace{0.45cm}

\begin{tabular}{|l c c c c r|}
  \hline
  \multicolumn{6}{|l|}{Input: $\z \in \Re^{128}$} \\
  \hline
  \hline
  \multicolumn{6}{|l|}{reshaped to (1, 1)$\times$128 channels} \\
  \hdashline[1pt/1pt]
  transconv &~1024~   &~~2$\times$1~~  &~~(2, 1)~~  &~BN~ &~ReLU \\
  transconv &256    &2$\times$1  &(2, 1)  &BN &ReLU \\
  transconv &256    &2$\times$1  &(2, 1)  &BN &ReLU \\
  transconv &256    &2$\times$1  &(2, 1)  &BN &ReLU \\
  transconv &128    &3$\times$1  &(3, 1)  &BN &ReLU \\
  \hdashline[1pt/1pt]
  transconv &64     &1$\times$7  &(1, 7)  &BN &ReLU \\
  transconv &$K_\b$ &1$\times$12 &(1, 12) &BN &tanh~ \\
  \hline
  \hline
  \multicolumn{6}{|l|}{Output: $G_\b\left(\z\right) \in \Re^{96 \times 84 \times K_\b}$ ($K_\b$-track piano-roll)} \\
  \hline
\end{tabular} \\
\vspace{0.15cm}
(b) the bar generator $G_\b$
\vspace{0.45cm}

\begin{tabular}{|l c c c r|}
  \hline
  \multicolumn{5}{|l|}{Input: $\widetilde{\x} \in \Re^{4 \times 96 \times 84 \times 5}$ (real/fake piano-rolls of 5 tracks)} \\
  \hline
  \hline
  \multicolumn{5}{|l|}{reshaped to (4, 96, 84) $\times$ 5 channels} \\
  \hdashline[1pt/1pt]
  conv~~~~~ &~~128~~ &~~$2\times1\times1$~~  &~~(1, 1, 1)~~  &~LReLU \\
  conv &128 &$3\times1\times1$  &(1, 1, 1)  &LReLU \\
  \hdashline[1pt/1pt]
  conv &128 &$1\times1\times12$ &(1, 1, 12) &LReLU \\
  conv &128 &$1\times1\times7$  &(1, 1, 7)  &LReLU \\
  \hdashline[1pt/1pt]
  conv &128 &$1\times2\times1$  &(1, 2, 1)  &LReLU \\
  conv &128 &$1\times2\times1$  &(1, 2, 1)  &LReLU \\
  conv &256 &$1\times4\times1$  &(1, 2, 1)  &LReLU \\
  conv &512 &$1\times3\times1$  &(1, 2, 1)  &LReLU \\
  \hdashline[1pt/1pt]
  \multicolumn{4}{|l}{fully-connected ~~~~1024} &LReLU \\
  \multicolumn{4}{|l}{fully-connected ~~~~~~\,1} & \\
  \hline
  \hline
  \multicolumn{5}{|l|}{Output: $D(\widetilde{\x}) \in \Re$} \\
  \hline
\end{tabular} \\
\vspace{0.15cm}
(c) the discriminator $D$
\vspace{0.45cm}

\begin{tabular}{|l c c c c r|}
  \hline
  \multicolumn{6}{|l|}{Input: $\y \in \Re^{96 \times 84}$ (piano-rolls of the given track)} \\
  \hline
  \hline
  conv~~~~\, &~~16~~ &~~$1\times12$~~ &~~(1, 12)~~ &BN &LReLU \\
  conv &16 &$1\times7$  &(1, 7)  &BN &LReLU \\
  \hdashline[1pt/1pt]
  conv &16 &$3\times1$  &(3, 1)  &BN &LReLU \\
  conv &16 &$2\times1$  &(2, 1)  &BN &LReLU \\
  conv &16 &$2\times1$  &(2, 1)  &BN &LReLU \\
  conv &16 &$2\times1$  &(2, 1)  &BN &LReLU \\
  \hline
  \hline
  \multicolumn{6}{|l|}{Output: $E(\y) \in \Re^{16}$} \\
  \hline
\end{tabular} \\
\vspace{0.15cm}
(d) the encoder $E$

\caption{Network architectures for the (a) temporal generator, (b) bar generator, (c) discriminator and (d) encoder. For convolutional (conv) and transposed convolutional (transconv) layers, the values represent (from left to right): number of filters, kernel size, strides, batch normalization (BN), and activation functions. For fully-connected layers, the values represent (from left to right): number of hidden nodes and activation functions. LReLU stands  for leaky ReLU.
$(K_\t, K_\b) = (1, 1), (1, 5), (5, 1)$ for the jamming, composer, and hybrid model, respectively.}
\label{tab:arch}
\end{table}

\newpage
\clearpage

\end{appendices}

\end{document}